# Wireless Sensor Network Virtualization: Early Architecture and Research Perspectives[±]

Imran Khan[+][¥], Fatna Belqasmi[#], Roch Glitho[*], Noel Crespi[+], Monique Morrow[++], Paul Polakos[++]

[+]Institut Mines-Télécom, Télécom SudParis, Evry, France
[#]Zayed University, Abu Dhabi, United Arab Emirates
[*]Concordia University, Montreal, Canada
[++]CISCO Systems
[¥]imran@ieee.org

## Abstract

Wireless sensor networks (WSNs) have become pervasive and are used in many applications and services. Usually the deployments of WSNs are task oriented and domain specific; thereby precluding re-use when other applications and services are contemplated. This inevitably leads to the proliferation of redundant WSN deployments. Virtualization is a technology that can aid in tackling this issue, as it enables the sharing of resources/infrastructure by multiple independent entities. In this paper we critically review the state of the art and propose a novel architecture for WSN virtualization. The proposed architecture has four layers (physical layer, virtual sensor layer, virtual sensor access layer and overlay layer) and relies on the constrained application protocol (CoAP). We illustrate its potential by using it in a scenario where a single WSN is shared by multiple applications; one of which is a fire monitoring application. We present the proof-of-concept prototype we have built along with the performance measurements, and discuss future research directions.

## Keywords

Wireless Sensor Network (WSN), Virtualization, Node-level virtualization, Network-level virtualization, Overlay Networks

## Introduction

In the last few years, Wireless Sensor Networks (WSNs) have become ubiquitous and are being used in a broad array of application domains, including healthcare, agriculture, surveillance and security. These WSNs are composed of small-scale nodes that have the ability to sense, compute and communicate [1]. While early sensor nodes were resource-constrained with limited capabilities, recent advances in sensor hardware technology have made it possible to produce sensor nodes that have more processing power, memory and prolonged battery life.

Virtualization is a key technique for the realization of the Future Internet, and it is indeed quite pertinent to explore it in the context of WSNs. Virtualization makes it possible to present physical computing resources by abstracting them into logical units, enabling their efficient usage by multiple independent users, including multiple concurrent applications [2]. Furthermore it even allows for the deployment of applications that were not even envisioned during an infrastructure's initial deployment.

To date, the realizations of WSNs have been domain-specific and task-oriented. Applications are bundled with a WSN at the time of deployment, and it is next to impossible to use the same WSN for another applications. This leads to redundant deployments and the underutilization of these resources. There are two approaches to allow multiple applications to access deployed WSN resources. One is to allow multiple applications to share the data gathered from a WSN. In this approach, a sink/gateway node collects all the data from the WSN and shares it among multiple users. For example, in [3], WSNs are merged into the

---

[±] This paper is an extended version of a short paper presented at "*6th Joint IFIP Wireless and Mobile Networking Conference (WMNC'13)*, April, 23-25, 2013, Dubai, UAE, under the title of "A Multi-Layer Architecture for Wireless Sensor Network Virtualization".

cloud by sending observed sensor data through a host manager that lies outside the WSN. The host manager simply collects the sensor data, profiles/aggregates it and then allows multiple applications to use it for their own purposes.

The second approach is to use the capabilities of the individual sensor nodes to execute multiple application tasks concurrently, and allow applications to group these sensor nodes together according to their requirements. The key difference between the two approaches is that the former approach allows the sharing of WSN data among multiple users, while the latter allows sharing of WSN nodes by multiple applications. This paper is focused on the second approach because it makes it possible to design more innovative applications over the deployed WSNs, even applications that were not envisioned a priori. This will greatly improve the efficiency of the deployed WSNs and will also encourage new business models.

This paper introduces the WSN virtualization concept, critically reviews the state-of-the-art in WSN virtualization and proposes a new early architecture which focuses on fixed WSNs. We illustrate the potential of the architecture by instantiating it for a fire monitoring scenario [4] in which multiple applications share the same WSN. We have built a prototype to demonstrate its feasibility and to measure its performance. We also identify further research directions.

The next section presents a critical overview of the state-of-the-art. The proposed architecture is presented in the third section. The fourth section discusses the implementation alternatives with the proof-of-concept prototype and the recorded performance measurements. The research directions are discussed in the fifth section. We conclude in the last section by discussing the lessons learned.

## 2. A Critical Overview of the State-of-the-Art

There are two categories of WSN virtualization: node level and network level. Figure 1 shows a high-level view of WSN virtualization. WSN node-level virtualization allows multiple applications to run their tasks concurrently on a single WSN node [5] (Fig. 1-a). This execution can be sequential (e.g. round-robin) or it can be simultaneous, with context switching between application tasks.

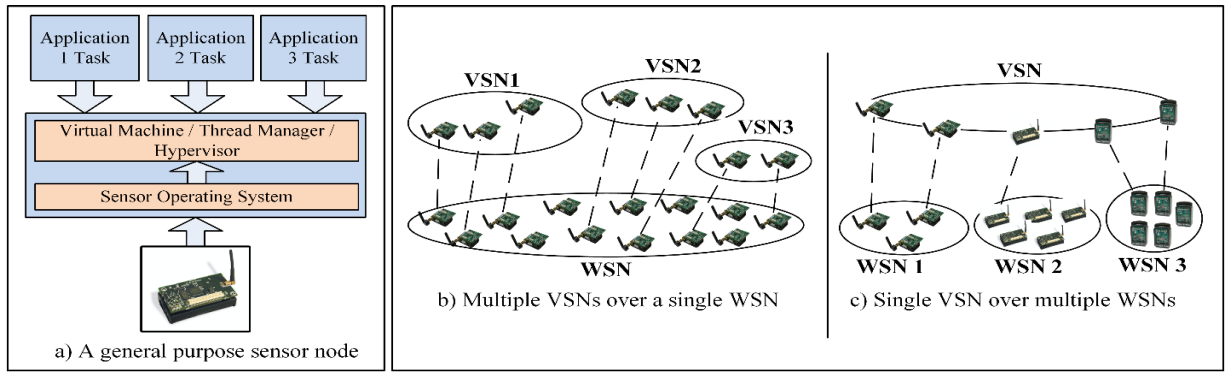

Figure 1: WSN Virtualization Categories

In WSN network-level virtualization, a subset of sensor nodes belonging to a deployed WSN form a Virtual Sensor Network (VSN) to execute given application tasks at a given time [6], while the other sensor nodes remain available for other application tasks. WSN network-level virtualization can be achieved in two ways. Different VSNs can be created over the same underlying WSN infrastructure (Fig. 1-b), or sensor nodes can form a single VSN over multiple WSNs in different administrative domains (Fig. 1-c). The latter situation is possible whenever the sensor nodes can support the concurrent execution of application tasks. It is the case these days because many popular sensor operating systems (e.g. Contiki) that run on resource constrained devices, enable node level virtualization through the concurrent execution of applications tasks on a same sensor node.

## 2.1. Motivating Example and Requirements

In this section we first present a motivating example and then draw requirements from it.

### 2.1.1. Motivating Example

A real-world deployment of a WSN is presented in [7], in which a WSN is used to monitor the impact of constructing a road tunnel under an ancient tower in Italy, as it was feared that the tower could lose its ability to stand on its own and that it might collapse during the construction. Now consider that there are three users interested in the fate of the tower. The first is the construction company, as it needs make sure that the tower does not lose its ability to stand on its own, otherwise it will have to pay a heavy fine. The second user is the conservation board which routinely monitors all the ancient sites around the city, and the third user is the local municipality which will have to plan emergency remedial/rescue actions in case the tower falls during the construction.

It is quite possible that the conservation board has already deployed its own WSN to monitor the health of ancient sites including this tower. In this case the construction company and the local municipality can reuse the existing sensor nodes during the construction period. In the absence of WSN virtualization, there are only two possible solutions. One is to rely on the information provided by the conservation board application. However this information may not be at the required granularity level. Worse, some of the information that is needed might simply not be available because the requirements of the construction company and of the local municipality were not considered when the conservation board application was designed and implemented. The second solution is that each user deploys redundant WSN nodes.

### 2.1.2. Requirements

The *first* requirement is the support of node-level virtualization to allow the execution of multiple application tasks on the same sensor node. The *second* requirement is the ability of sensor nodes to dynamically form groups to execute isolated and transparent application tasks concurrently (i.e. support for WSN network-level virtualization). The *third* requirement is the support of application priority. In some critical application scenarios such as fire monitoring, it is important that other tasks have less priority than the one reporting the fire event.

The *fourth* requirement is that the proposed solution should be applicable to a wide range of applications and should not be tailored for a particular scenario or domain, which is the usual case with most solutions. The *fifth* requirement is that the proposed solution should be platform-independent and should not depend on specific operating systems or customized/tailored interfaces. The *sixth* and final requirement is that the solution should address heterogeneity, i.e., cope with sensor nodes that have different capabilities (e.g. processing power, memory).

## 2.2. The State-of-the-Art and its Shortcomings

We divide the related work into three classes: node-level, network-level and hybrid virtualization solutions. The hybrid solutions combine both node- and network-level virtualization.

### 2.2.1. Node-level Virtualization

In order to achieve node-level virtualization, mechanisms must be in place to allow deployed WSN nodes to execute new application tasks as well as update existing ones. One solution is to reprogram WSN nodes individually, but that is neither feasible nor efficient. Wireless reprogramming, on the other hand, allows large number of WSN nodes to be updated with new application tasks with minimum effort. It is now the main mechanism used for node-level virtualization. Two examples of node-level virtualization based on wireless reprogramming are discussed below. Their main drawback is platform dependency.

Maté [5] is a pioneering work that provides sequential execution of application tasks on resource-limited, early generation sensor nodes. It is a tiny virtual machine consisting of a stack-based binary code interpreter and works on top of TinyOS. Application tasks are divided into code capsule(s) of up to 24 instructions and are executed one by one. A viral code distribution scheme is used to propagate code and to reprogram the sensor nodes. As there is tight coupling between the application code and TinyOS, installing a new code requires the replacement of the whole OS. There is no support for application priority, and only a limited set of applications is supported. Furthermore, the approach is not platform-independent since it only works on TinyOS, but it does address heterogeneity.

MANTIS [8] is a thread-based embedded operating system. Programs are created as user-level threads with dedicated memory space and static data attached to them at compile time. Long-running threads can be pre-empted by short-running threads. The work on wireless reprogramming is ongoing, according to the authors. The techniques being used are the wireless re-flashing of the OS and the re-programming of single threads. Unlike Mate, MANTIS does provide application priority. However, it is not platform-independent.

### 2.2.2. Network-level Virtualization

In [6], sensor nodes form clusters to support applications that monitor dynamic phenomena. The sensor nodes within each cluster executes application(s) tasks, meaning a sensor node can be part of multiple clusters. With each cluster dedicated to an application, a WSN can be utilized by multiple applications concurrently, hence realizing network-level virtualization. Two illustrative applications are presented as motivation. Unfortunately the work is poor in terms of technical details (e.g. how individual nodes execute application tasks). Furthermore, there is no discussion of how application priority, heterogeneity and platform independence are tackled. This work has been extended in reference [9] in order to facilitate the creation, operation and maintenance of dynamic clusters to achieve network-level virtualization. Once an event is detected, sensor nodes are grouped as a dynamic cluster tree by exchanging VSN formation messages. However, in terms of our requirements none of the drawbacks of reference [6] are addressed.

The authors in [10] introduce the problem of mission assignment in WSNs. The work can be related to network-level virtualization because the WSN is able to support multiple missions at the same time. Each mission uses a dedicated subset of sensor nodes which are not shared with other missions. A mission assignment problem is modelled as a weighted bipartite graph to optimally assign the sensor nodes to missions. Achieving a mission produces a profit, so the goal is to maximize profit by efficiently achieving as much missions as possible. Both centralized and distributed solutions are presented, using proofs and algorithms including an energy aware solution. This solution does not consider any specific application domain. Heterogeneity is addressed along with platform independence. However, application task priority is not provided since each sensor node executes only one application task at a time.

### 2.2.3. Hybrid Solutions

The authors in [11] discuss the SenShare platform, which supports both WSN-node and network-level virtualization. They consider TinyOS applications with an embedded hardware abstraction layer (HAL). The underlying sensor node resources are then accessed using a run-time layer on top of TinyOS. Since TinyOS supports multiple tasks at the same time, node-level virtualization is thus achieved. For network-level virtualization, an overlay network using Collection Tree Protocol (CTP) is created to group sensor nodes executing the same application. The physically scattered sensor nodes executing the same application can be grouped into a single overlay network. SenShare is the first solution targeting comprehensive WSN virtualization. It supports node- and network-level virtualization, application priority and heterogeneity, and it is independent of any application domain. However, it is not platform-independent, as only TinyOS applications are supported.

Melete [12] is an extension of Maté and supports both node- and network-level virtualization. Concurrent execution of application tasks is achieved by making the following enhancements to Maté: dedicated storage and execution space for applications to allow concurrency, and a code dissemination protocol to allow selective and reactive (re)programming of sensor nodes. For network-level virtualization it uses a dynamic grouping technique of sensor nodes. A sensor node can be part of more than one logical group at the same time. The supported network topology is a connected graph. Melete does not support application priority, and is not platform-independent. It only supports a limited set of applications, but it does tackle heterogeneity.

## 3. Proposed Architecture

In this section, we first present the architectural principles. We then present our multi-layer architecture based on overlays, followed by a discussion of the interfaces and the overlay creation procedure.

### 3.1 Architectural Principles

The first architectural principle is that new applications/services are deployed as new overlays on top of the physical WSN. Overlays have several advantages: they are distributed, lack central control and allow resource sharing [13]. The second principle is that any given physical sensor node can execute (locally) a task for a given application deployed in the overlay. Any given sensor node may execute several such application tasks at any given time.

The third principle is that not all WSN nodes perform the overlay-related operations, as they may not have enough capabilities to support the overlay middleware. When that is the case, they will delegate the operations to more powerful sensors and even to other nodes. This principle in effect makes it possible to address the heterogeneity requirement and enables network level virtualization for current generation resource constrained sensor nodes.

The fourth principle is that within the architecture there are separate data and control paths. The sensor data (e.g. temperature values) is transmitted from sensor nodes to the overlay application using the data path. The control data (e.g. changing application priority and overlay management) is sent over the control path. This separation of paths makes it easy to work on new protocols for each path independently.

The last principle is the use of emerging standards, aimed at resource-constrained devices, to tackle the platform independence challenge. These standards include protocols such as the Constrained Application Protocol (CoAP) [14], DNS-Service Discovery (DNS-SD) [15] and standards such as Sensor Model Language (SensorML) [16], Observations & Measurements (O&M) [17] and Sensor Markup Language (SenML) [18]. This principle of course implies the need for converters/mappers for devices which do not support the standards.

CoAP is an application layer transfer protocol, like HTTP, designed to work with resource-constrained devices. It has less overhead, memory and processing requirements than HTTP. DNS-SD offers service discovery in resource-constrained networks and allows for the seamless integration of such architectures to the existing IP networks. SensorML provides standard models and XML-based encoding to describe sensor measurements and processes. It is able to provide interoperability, automatic discovery, utilization and sensor sharing. O&M is a standard which defines encoding schemas for the observations made by sensors. SenML provides a data model for sensor measurements and simple metadata about sensors in JSON, XML and EXI formats.

## 3.2 Overall Architecture

Figure 2 shows our proposed multi-layer architecture, and Table I provides the list of components used. There are four layers (physical, virtual sensor, virtual sensor access and overlay), two paths (data and control), five interfaces (data ($D_i$), proprietary D$i$ ($PD_i$), control ($C_i$), proprietary C$i$ ($PC_i$) and gateway ($G_i$)) and a registration server.

TABLE I. COMPONENTS OF THE ARCHITECTURE

| Abbreviation | Component | Remarks |
|---|---|---|
| – | Type A Sensor | Legacy/resource constrained sensor |
| – | Type B Sensor | New generation smart IP sensor node |
| GTO Node | Gates-to-Overlay Node | Gateway/sink node capable of joining application overlays on behalf of Type A sensors |
| – | Sensor Agent | Functional entity providing a unified interface to provide platform independence |
| – | Registration Server | Sensor repository |
| *Di* | Data Interface | Interface to send sensor data to application overlay |
| *PDi* | Proprietary Data Interface | Proprietary interface to send virtual sensor data to sensor agent |
| *Ci* | Control Interface | Interface to send/receive control data from end-user application |
| *PCi* | Proprietary Control Interface | Proprietary interface to send/receive control data from virtual sensor to sensor agent |
| *Gi* | Gates-to-Overlay Interface | Interface to send/receive the control data between Type A sensors and Type B sensors/GTO nodes |

At the physical layer we have independent WSNs that consist of two types of sensor nodes, i.e., resource constrained (type A) and capable (type B) sensors. Each WSN also has specialized nodes, called GTO nodes. Their role is to help type A sensors join the application overlays and provide heterogeneity. Gateways, sink nodes or a type B sensors can act as GTO nodes when required. For example, in the motivating example in section 2.1.1, if the existing sensors are of type A, then either the existing gateway node or Type B sensors, deployed by the construction company, can help those sensors to become part of the construction company overlay. This might increase the complexity of the type B sensor nodes but it does allow flexibility.

The virtual sensor layer consists of the logical representation of each sensor executing multiple application tasks concurrently. Each logical representation is called a virtual sensor in our architecture, which is an abstraction of an application task run by a sensor.

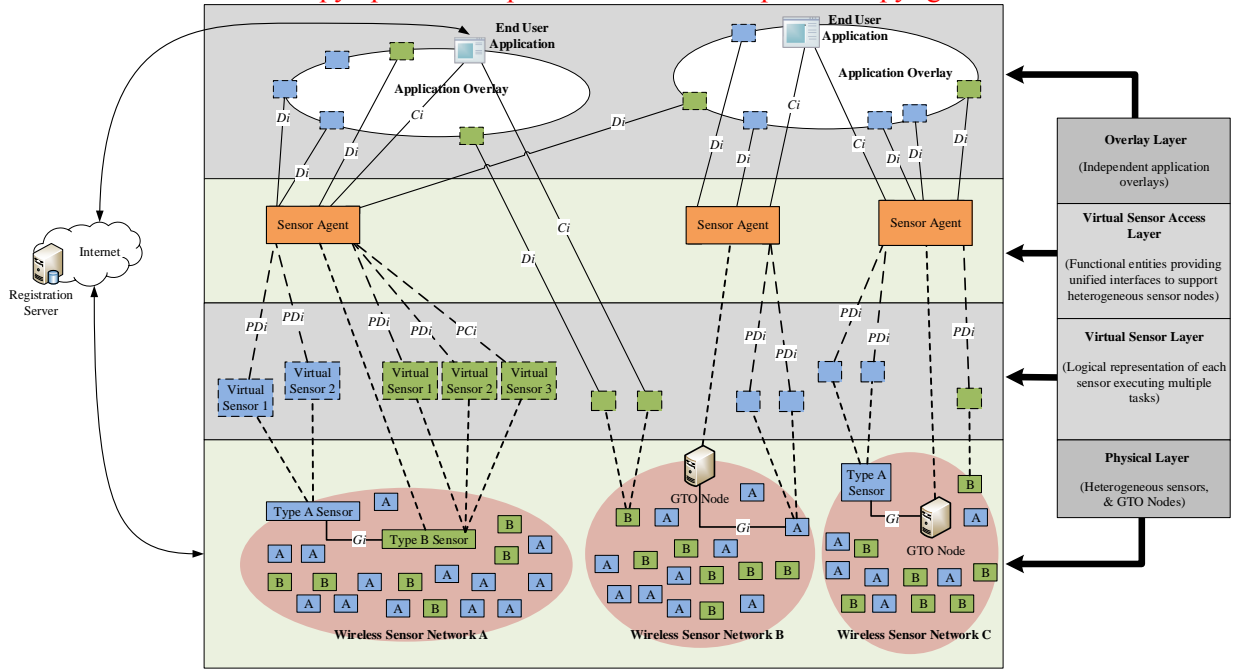

**Figure 2: Multi-layer WSN Virtualization Architecture**

The virtual sensor access layer consists of sensor agents which ensure platform independence. This is achieved by providing standardized interfaces (*Di* and *Ci*) to interact with the end-user applications, and are mapped onto platform-specific (proprietary) interfaces (*PDi* and *PCi*) for the underlying physical sensor nodes. Sensor agents can be implemented either in capable (type B) sensors or in GTO nodes.

The overlay layer consists of independent application-specific overlays (two are shown in the figure 2, but there could be many more). Each application overlay is created by the end user application and consists of virtual sensors that run the overlay application tasks. An overlay protocol is used for message exchange inside an overlay. A Registration Server, which contains the details of the deployed sensor nodes, is used by end-user applications to find sensor nodes.

### 3.3 Interfaces

The data path uses the data interface (*Di*) supported by all of the sensor agents to send the data received from the virtual sensors executing the end user's application task to the application overlays. The control path uses the control interface (*Ci*) supported by all sensor agents to send/receive control data. Examples of control data include sending requests to change application priority and sampling frequency. The interfaces, *PDi* and *PCi* are proprietary interfaces and are used by the sensor agent to communicate with WSNs. Figure 3 shows high level examples of when sensor data is sent over *PDi* and *Di* interfaces (3a) (when fire is detected) and when a request to change application task priority is sent over *Ci* and *PCi* interfaces (3b). In this case it is the priority of the task running on sensor 02. The Gates-to-overlay interface (*Gi*) is provided by all the sensors as well as the GTO nodes. Any communication from type B or GTO nodes with type A sensors is done using this interface.

### 3.4 Overlay Creation Procedure

This section describes the overlay creation procedure. The creation of the overlay is a three step procedure, initiated by the end user application. The first step is the dynamic resource discovery and overlay pre-configuration, allowing the discovery of the sensors and GTO nodes on the fly according to the requirements of the end user application. The second step is the activation of the overlay. The selected sensor (type B) and GTO nodes receive an overlay join request (or advertisement) over the *Ci* interface. After joining the overlay, the type B sensors and the GTO nodes (for type A sensors) may receive the application task, with its desired priority level. The final step is the execution of the end user application, which begins when each sensor starts executing the end user application task. Depending on the application requirements, sensors may exchange messages among themselves in the overlay before sending any data to the end user application over the *Di* interface.

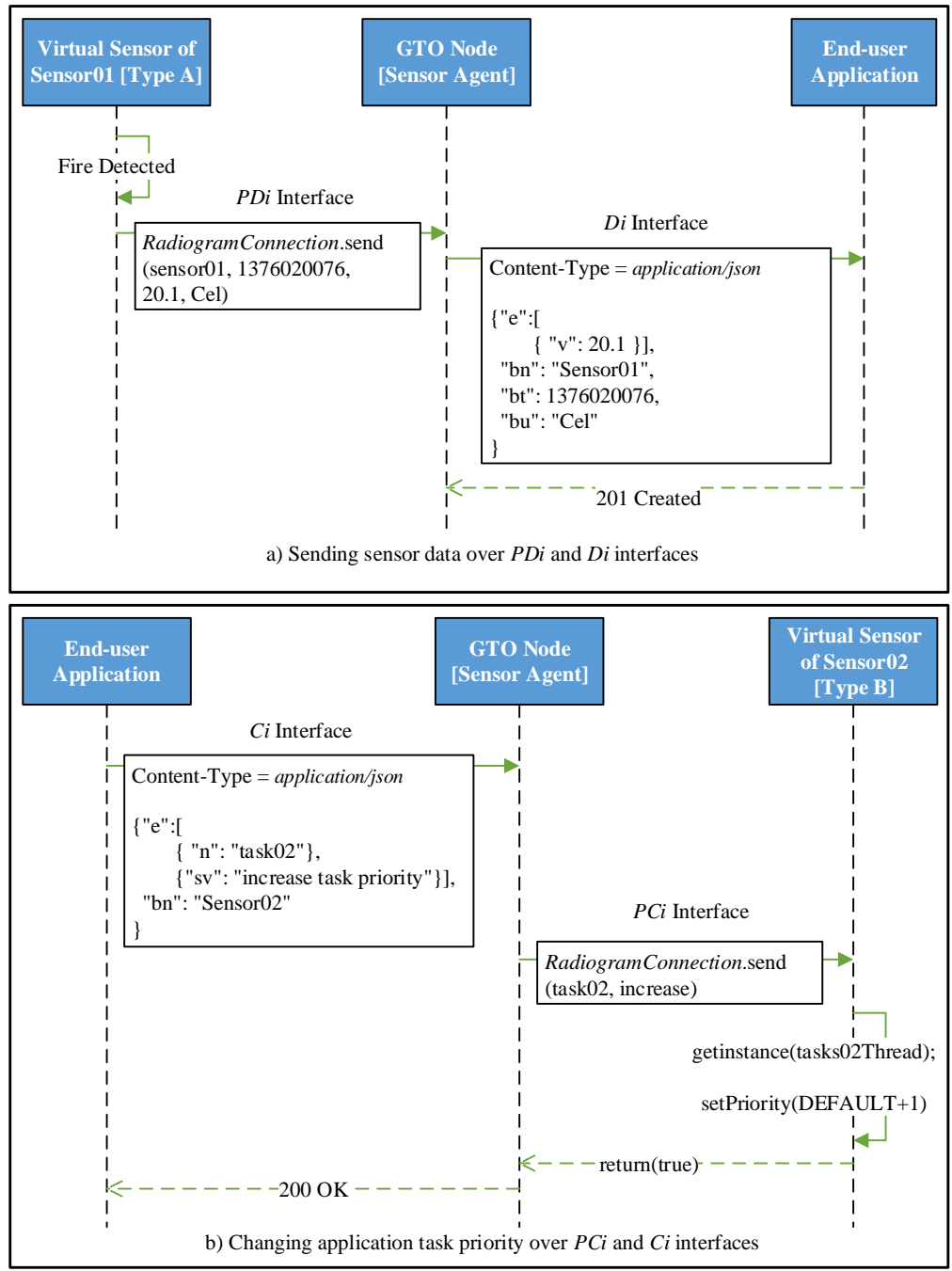

Figure 3: **Example of communication over data and control interfaces**

# 4 Implementation Alternatives, Proof of concept Prototype and Measurements

## 4.1 Implementation Alternatives

Our proposed architecture consists of the data plane, the control plane and several interfaces that belong to them. The *Di* interface, belonging to the data plane, carries the actual data. The *Ci* and *Gi* interfaces carry control messages and are part of the control plane.

There are several options for implementing a data plane interface. Both HTTP and CoAP can be used as application layer protocols, but we chose CoAP as it will allow type A nodes to support the same protocol for *Di* and *Gi* interfaces. We use SenML specifications to encode the sensor data in standard JSON format. The combination of SensorML and O&M is another option, but we selected SenML since it is less complex.

For the control plane, one candidate protocol is JXTA [19], an open source peer-to-peer protocol specification that allows the creation of independent, robust and efficient overlay networks. ScatterPastry [20] is another option. For our work we opted to use JXTA since its implementations are readily available.

## 4.2 Prototype

We implemented a simple brush fire scenario discussed in [4] as a prototype. In this scenario, the city administration is interested in the early detection of brush fire eruption and in its evolution, using a WSN and a Fire Contour Algorithm (FCA). Some houses in the area already have their own sensors to detect fire. To accelerate the deployment of its application and to avoid redundancy, the city administration has opted to deploy sensors in areas under its jurisdiction (i.e. streets and parks) and to incorporate the sensor nodes already deployed in private homes. The home owners get incentives like tax rebates for allowing the use of their sensors by city administration. The home gateways acts as GTO nodes. All of the privately-owned sensors execute two application tasks – one for the home owner and one for the city administration. Figure 4a shows the mapping of the scenario onto our architecture.

We make the following assumptions. First we assume that the city administration has already discovered and sent its application task to each of these sensors. The second assumption is that all of the sensors in the prototype are type A sensors which need a GTO node for overlay-related tasks. Third, as it was not possible to generate a fire in a lab environment, the city administration application task periodically measured the temperature value in a sensor and sent it to the GTO node. We used six Java SunSpots sensors, each executing three application tasks concurrently. The application tasks and the FCA were coded in Java 2 Platform Micro Edition (J2ME). J2ME is a robust, flexible Java platform that enables the development of applications for mobile and embedded devices. The city administration's overlay network was implemented using a Java based implementation of the JXTA protocol, JXSE 2.6.

A RESTful web service is used by the city administration node to receive fire alerts. Each GTO node, upon receiving fire notification from its sensor, sends an HTTP POST message to a URI (*http://.../FireContourService/events/fire/*) to create a fire event. The content type of the HTTP POST message is set to *application/senml+json* and the event data received from Java SunSpot is mapped to JSON format according to SenML specifications. Once the event is created, the city administration node sends a fire notification message to the peers in the overlay.

The overlay is created by the city admin node, acting as rendezvous peer, by advertising its peer group (*fire contour service*) using JXTA pipe advertisements before the fire event. The GTO nodes join the fire contour service as edge peers by replying to the received pipe advertisement. The city admin node sends the fire notification message using the JXTA multicast socket, which provides efficient message exchange between members of the same peer group. After the execution of the fire contour algorithm, the reply message is sent directly to the city admin node instead of being multicast.

The prototype uses a simple probabilistic fire contour algorithm, considering that a distant house will send fire notifications less frequently than a nearby house because the fire is far from it. The city administrations' application is created using JavaFX, and receives the fire alert messages as well as the peers' replies and displays the output on the area map. JavaFX is a set of Java libraries that allow developers to rapidly design, create and deploy client applications that operate across diverse platforms.

The prototype setup is illustrated in Fig. 4b. The city administration application and its fire contour web service ran on a laptop with an Intel Core i5 CPU clocked at 2.67 GHz, and a 4GB RAM with 32bit Windows 7 Enterprise. The other two laptops acted as GTO nodes for Java SunSpots and ran three JXTA peers each. Their configurations were an Intel Core i7 CPU clocked at 2.70 GHz with 8GB RAM, 64-bit Windows 7 Professional and an Intel Core i5 CPU clocked at 2.60 GHz, and a 4GB RAM with Windows 7 Enterprise. All three laptops used JVM version 1.7.0_21 and were connected to a private LAN.

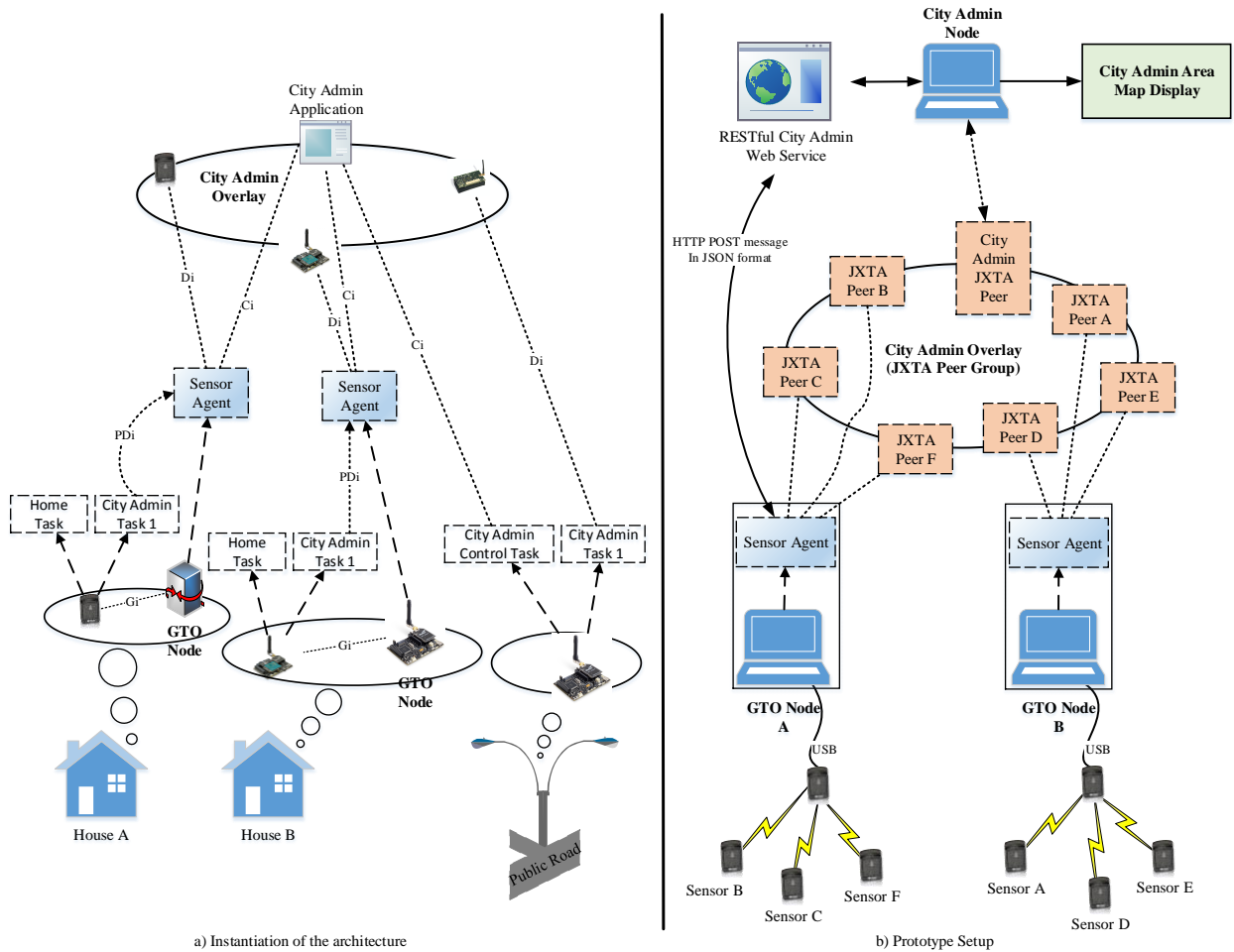

**Figure 4: Instantiation of the architecture and Prototype setup**

### 4.3 Performance Measurements

***Performance Metrics*** – The performance of the prototype was assessed in terms of the following delays: *HTTP POST Delay* (HPD), *Overlay Creation Delay* (OCD) and *Fire Notification Delay* (FND).

HPD is the time difference between when the GTO node sends an HTTP POST request and when it receives the corresponding success code (201 created). HPD is calculated for each sensor. OCD is the time it takes to set up the city administration overlay from a non-existent state to a ready state, when it advertises its fire contour service and is ready to accept join requests. We measured this delay inside the Java code to ensure that the OCD does not include the JVM start-up delay. FND is measured as the time it takes for the city admin node to multicast fire notification messages to JXTA peers and to receive their replies after they execute fire contour algorithm. For each experiment we restarted the JVM and cleared the previous JXTA configuration cache. All delays are measured in milliseconds and calculated at the sender side.

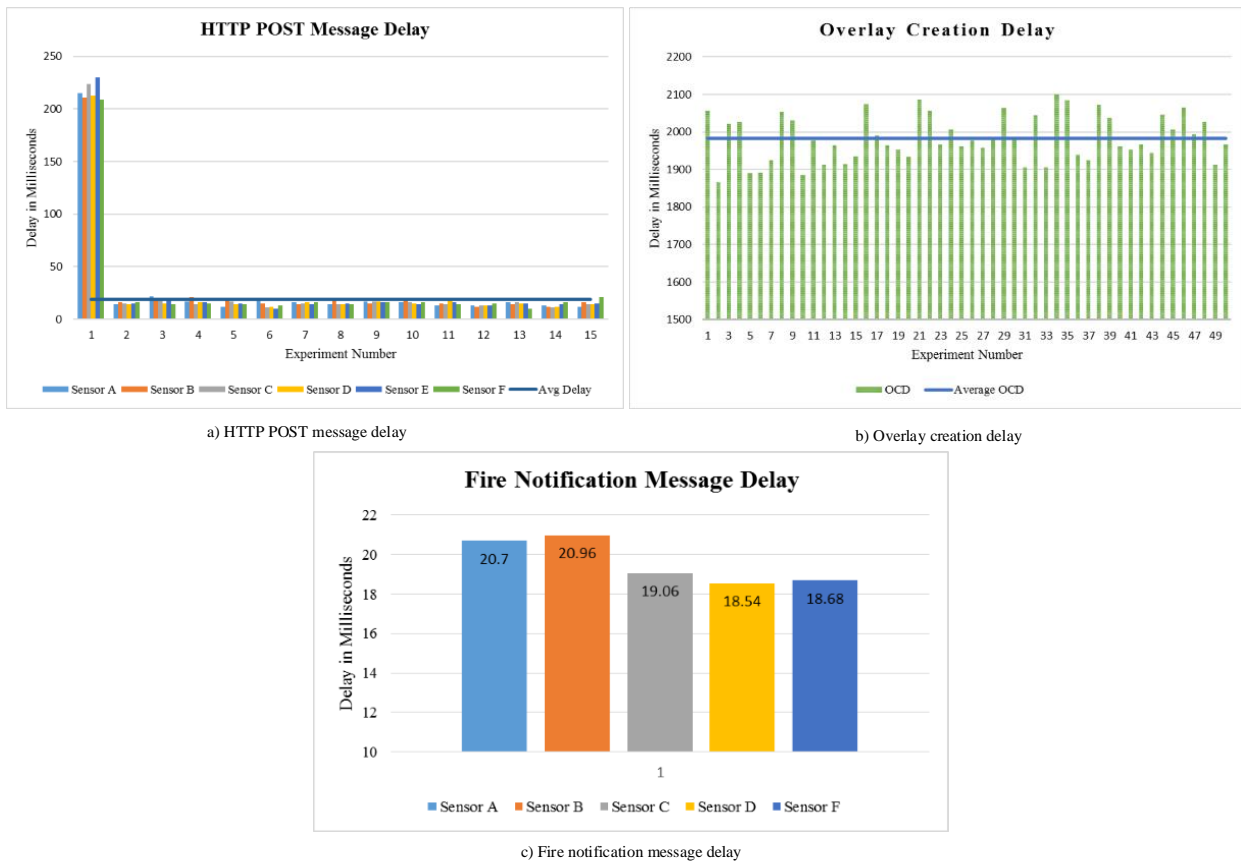

a) HTTP POST message delay  
b) Overlay creation delay  
c) Fire notification message delay

**Figure 5: Results**

***Performance Results*** – The HPD measurements are shown in Fig. 5(a) (for clarity, only 15 measurements are shown). The dark blue horizontal line shows the average delay for the *50* measurements, *18.96ms*. It is observed that the delay for first POST message is much larger than that for the subsequent messages. This long delay is due to the three-way handshake of TCP connection that takes place during the first POST message, whereas for subsequent requests a persistent HTTP connection (a.k.a. HTTP keep-alive) reduces delay considerably. Figure 5(b) shows the OCD of a city admin JXTA peer with an average value of *1983ms* from *50* iterations indicated by the horizontal blue line. The delay includes the JXTA core start-up, the creation of a fire contour service, related pipe advertisement, a JXTA multicast socket and the thread for accepting join requests from other JXTA peers. For each iteration a new JXTA cache was generated instead of using the old one. Figure 5(c) shows the average FND of five sensors that executed a fire contour algorithm in response to the notification message sent by a city admin JXTA peer. In this case sensor E reported the fire. The average FND of five sensors is *19.58ms*.

In order to determine the overhead of WSN virtualization, we consider the scenario where sensors do not support node-level virtualization and only execute city admin tasks. There is also no network-level virtualization and no overlay network for message exchange. In this case, the fire counter algorithm will be executed by the GTO nodes after getting an HTTP POST message from the city admin node. For a simple comparison, if we consider that the FND without WSN virtualization is similar to HPD, i.e., *18.96ms*, and FND with WSN virtualization is *19.58ms*, then with WSN virtualization we have approximately *3.27%* overhead. This overhead is due to the processing of XML-based JXTA messages. Our implementation demonstrates that WSN virtualization is indeed feasible and does not incur much overhead. Node-level virtualization is achieved with Java SunSpots with very little effort. Network-level virtualization is achieved using JXTA, and once JXTA is operational, the delays are minimal. OCD is inevitable, but in the long-run, using JXTA is beneficial as it provides a robust, highly scalable and efficient solution.

Overall the results show the typical delays experienced in a private LAN setting. The same JXTA pipe advertisement of the fire contour service was used to send and receive the fire notification messages over JXTA multicast socket, which greatly improved the overall performance.

## 5 Research Directions

WSN virtualization is a very rich research area and our proposed preliminary architecture has raised several interesting issues. This section provides a non-exhaustive sample. A first issue is a dynamic publication and discovery framework for sensor and GTO nodes. In this work, we assumed a static publication process where the sensor and GTO owners publish their nodes to a central repository. To automate the process of WSN virtualization, an on-the-fly publication and discovery mechanism would be required. A CoAP-based framework could be used as starting point. For a centralized solution, a CoAP Resource Directory (RD) mechanism can be used, while a CoAP resource discovery mechanism would be more appropriate for a distributed solution. Similarly, a DNS-SD mechanism can be used in combination with CoAP to provide new, powerful solutions.

The choice of data formats for various interfaces is another issue. The current OGC – O&M and SensorML specifications use the XML format, which is inefficient in resource-constrained environments. SenML addresses this issue by using JSON and EXI formats, and it works with both HTTP and CoAP, but it also has some open issues. For example, we can use it to specify simple metadata about measurements but there is no mechanism to provide such data for describing the sensors, their capabilities and their resources (memory, space, and battery-life) at a particular time. The possibility of a lightweight mechanism for reporting a sensors' run-time status is very appealing. Similarly, a semantically-enriched format would be of particular use for creating intelligent sensor-based systems in the context of IoT, which is currently not possible with SenML.

An important issue is optimal task assignment to sensors. The problem is essentially the mapping of end-user application requirements to the available resources, which is very challenging in a virtualized environment. Reference [10] proposes a solution, but it assumes that every sensor executes a single task, which is not the case in a virtualized environment. However, it could be used as starting point for further research. WSN-oriented overlay middleware is yet another issue to investigate. We need an efficient solution that prevents overlays from interacting in a harmful way when they compete for underlying resources. JXTA and similar protocols work well, but not in resource-constrained environments. Some early attempts like [20] exist, but they must be combined with the concept of WSN virtualization.

A signalling framework to support Quality of Service (QoS) and session management is also needed. Issues like handling application requests for setting/changing task priority will be tackled by such a general QoS framework. There are several signalling frameworks, such as SIP/RSVP, but they may not be suitable for sensors. Again, a CoAP-based signalling protocol is a potential solution. Virtualization as applied to mobile WSNs is also a key issue, since mobile WSNs are becoming more and more popular. Vehicular ad hoc networks, social networks and crowd-based sensing can provide concrete application scenarios to motivate the virtualization of mobile WSNs.

## 6  Lessons learned

In this paper we have proposed a new preliminary multi-layer architecture for WSN virtualization and have identified several research directions.

We have learned several lessons. The first is that WSN node-level virtualization is still in its infancy and very few WSN kits supporting node level virtualization are readily available. This is certainly due to the challenges of designing hypervisors in resource-constrained environments. A second lesson is that most existing WSN standard specifications pertinent to our work are still embryonic. SenML, for instance, is very promising. However, in its present form, it is not suitable for control functions. On the other hand, SensorML is complex and comes with additional functionalities that are unsuitable for a general purpose and efficient solution. A third lesson is that most existing overlay middleware are unsuitable for WSN because they are usually not designed for resource-constrained devices. We used JXSE, which is one of the best choices available. However, its current open source implementation is rather old and the future of the initiative is uncertain.

### Acknowledgement

This work is partially supported by CISCO systems through grant CG-576719, and by the Canadian Natural Science and Engineering Research Council (NSERC) through the Canada Research Chair in End-User Service Engineering for Communications Networks.

**Biographies**

IMRAN KHAN (imran@ieee.org) received his BCS degree in Computer Science from COMSATS Institute of IT, Pakistan in 2005 and M.S. degree in Multimedia and Communication from M.A. Jinnah University, Pakistan in 2009. Since May 2011 he is a Ph.D. research student in CNRS Lab UMR5157 at Institut Mines-Télécom, Télécom SudParis jointly with Paris VI (UPMC). Currently he is a collaborating researcher at Concordia University, Montreal working on a CISCO project. In past he worked on projects funded by ISOC and ITEA2. He is student member of IEEE. His research interests are Virtualization, Wireless Sensor Networks, Internet of Thing (IoT), and M2M Communications.

FATNA BELQASMI (fatna.belqasmi@zu.ac.ae) holds a Ph.D. and an M.Sc. degree in electrical and computer engineering from Concordia University, Canada. She is current working as Assistant Professor at Zayed University Abu Dhabi, UAE. In the past, she worked as a research associate at Concordia University, Canada and as a researcher at Ericsson Canada. She was part of the IST Ambient Network project (a research project sponsored by the European Commission within the Sixth Framework Programme -FP6-). She worked as an R&D engineer for Maroc Telecom in Morocco. Her research interests include next generation networks, service engineering, distributed systems, and networking technologies for emerging economies.

ROCH GLITHO [SM] (http://users.encs.concordia.ca/~glitho/) holds a Ph.D. (Tekn. Dr.) in tele-informatics (Royal Institute of Technology, Stockholm, Sweden) and M.Sc. degrees in business economics (University of Grenoble, France), pure mathematics (University Geneva, Switzerland), and computer science (University of Geneva). He works in Montreal, Canada, as an associate professor of networking and telecommunications at the Concordia Institute of Information Systems Engineering (CIISE) where he leads the telecommunication service engineering (TSE) research laboratory (.http://users.encs.concordia.ca/~tse/). In the past he has worked in industry for almost a quarter of a century and has held several senior technical positions at LM Ericsson in Sweden and Canada (e.g. expert, principal engineer, senior specialist). His industrial experience includes research, international standards setting (e.g. contributions to ITU-T, ETSI, TMF, ANSI, TIA, and 3GPP), product management, project management, systems engineering and software/firmware design. In the past he has served as IEEE Communications Society distinguished lecturer, Editor-In-Chief of IEEE Communications Magazine and Editor-In-Chief of IEEE Communications Surveys & Tutorials. His research areas are: virtualization and cloud computing; Machine-to-Machine communications (M2M) and Internet of Things; Distributed systems (e.g. SOAP Based – Web Services, RESTful Web Services); Rural communications and networking technologies for emerging economies.

NOEL CRESPI (noel.crespi@mines-telecom.fr) holds a Master's from the Universities of Orsay and Kent, a diplome d'ingénieur from Telecom ParisTech, and a Ph.D. and a Habilitation from Paris VI University. He worked from 1993 in CLIP, Bouygues Telecom, France Telecom R&D in 1995, and Nortel Networks in 1999. He joined Institut Mines-Télécom in 2002 and is currently professor and program director, leading the Service Architecture Laboratory. He is appointed as coordinator for the standardization activities in ETSI and 3GPP. He is also a visiting professor at the Asian Institute of Technology and is on the four-person Scientific Advisory Board of FTW, Austria. His current research interests are in service architectures, P2P service overlays, future Internet, and Web-NGN convergence. He is the author/co-author of more than 230 papers and contributions in standardization.

MONIQUE MORROW (mmorrow@cisco.com) holds the title of CTO Cisco Services. Ms. Morrow's focus is in developing strategic technology and business architectures for Cisco customers and partners. With over 13 years at Cisco, Monique has made significant contributions in a wide range of roles, from Customer Advocacy to Corporate Consulting Engineering. With particular emphasis on the Service Provider segment, her experience includes roles in the field (Asia-Pacific) where she undertook the goal of building a strong technology team, as well as identifying and grooming a successor to assure a smooth transition and continued excellence. Monique has consistently shown her talent for forward thinking and risk taking in exploring market opportunities for Cisco. She was an early visionary in the realm of MPLS as a technology service enabler, and she was one of the leaders in developing new business opportunities for Cisco in the Service Provider segment, SP NGN. Monique holds 3 patents, and has an additional nine patent submissions filed with US Patent Office. Ms. Morrow is the co-author of several books, and has authored numerous articles. She also maintains several technology blogs, and is a major contributor to Cisco's Technology Radar, having achieved Gold Medalist Hall of Fame status for her contributions. Monique is also very active in industry associations. She is a new member of the Strategic Advisory Board for the School of Computer Science at North Carolina State University. Monique is particularly passionate about Girls in ICT and has been active at the ITU on this topic - presenting at the EU Parliament in April of 2013 as an advocate for Cisco. Within the Office of the CTO, first as an individual contributor, and now as CTO, she has built a strong leadership team, and she continues to drive Cisco's globalization and country strategies.

PAUL POLAKOS (ppolakos@cisco.com) is currently a Cisco Fellow and member of the Mobility CTO team at Cisco Systems focusing on emerging technologies for future Mobility systems. Prior to joining Cisco, Paul was Senior Director of Wireless Networking Research at Bell Labs, Alcatel-Lucent in Murray Hill, NJ and Paris, France. During his 28 years at Bell Labs he worked on a broad variety of topics in Physics and in Wireless Networking Research including the flat-IP cellular network architecture, the Base Station Router, femtocells, intelligent antennas and MIMO, radio and modem algorithms and ASICSs, autonomic networks and dynamic network optimization. Prior to joining Bell Labs, he was a member of the research staff at the Max-Planck Institute for Physics and Astrophysics (Munich) and visiting scientist at CERN and Fermilab. He holds BS, MS, and Ph.D. degrees in Physics from Rensselaer Polytechnic Institute and the University of Arizona, is a Bell Labs and Cisco Fellow, and author of more than 50 publications and 30 patents.